\documentclass{article}
\usepackage{amsmath}
\usepackage{amsfonts, amssymb}%
\usepackage{braket}
\usepackage{mathtools}
\usepackage[margin=1in]{geometry}
\usepackage[caption=false]{subfig}
\usepackage{graphicx}
\usepackage[dvipsnames]{xcolor}
\usepackage[normalem]{ulem}
\usepackage{authblk}
\usepackage{adjustbox}
\usepackage{tensor}
\usepackage{bbm}
\usepackage{relsize}
\usepackage[sort&compress,numbers]{natbib}
\usepackage{appendix}

\usepackage[citebordercolor=blue]{hyperref}
\hypersetup{colorlinks,linkcolor={blue},citecolor={purple},urlcolor={blue}}
\usepackage[capitalise]{cleveref}

\title{Linear multi-particle, multi-port interferometers and their matrix representations}
\author{B\"ulent Demirel}

\affil[1]{University of Stuttgart,
	Institute for Functional Matter and Quantum Technologies,
	Allmandring 3,
	D-70569 Stuttgart}
\affil[1]{ buelent.demirel@fmq.uni-stuttgart.de}
\crefname{app}{Appendix}{Appendices}

\begin{document}
\maketitle

\begin{abstract}
Interferences in multi-path systems for single and multiple particles are theoretically analyzed. A holistic method is presented, which allows to construct the unitary transition matrix describing interferometers for any port number $d$ and particle number $n$. This work has two centerpieces. First, a solution algorithm for realizing any desired unitary transformation in an interferometer for the single particle case is demonstrated. Secondly, a simple scheme for calculating  the rotation matrices of beam splitters and phase shifters for multiple particles is shown. Such linear optical devices are especially important for quantum circuits and quantum information processing. 

\end{abstract}

\section{Introduction}

Particle interference is an essential wave phenomenon for researches in quantum information theory in the field of quantum optics. Multi-port interferometry (MPI) is considered as a potential contestant for practical realizations of quantum computation \cite{Knill2001, Kok2007, Howell2000} and algorithms~\cite{Tabia2016, Barak2007}, especially with photonic circuits on quantum chips~\cite{Berry2009, Peruzzo2011, Crespi2013, Carolan2015}. One of the most obvious applications of its kind is boson sampling~\cite{Aaronson2013, Tichy2014, Clifford2018}. 
Many fundamental quantum effects are tested in interferometers, often of the Mach-Zehnder type, with all kinds of particles~\cite{mandel1999, rauch2015, berman1997, hornberger2012}.\\

Single particle multiport interferometers, as examined by Reck~et~al.~\cite{Reck1994}, provide a way of realizing any finite unitary operator by using spatial modes of linear optical elements. 
A convenient property of this mathematical description is that the logic of mode mixing is taken care of by standard matrix multiplication. Naturally, it is desirable to have such a description for a many-particle system as well. \\

In this letter new ways to generalize the existing concept are provided and demonstrated along concrete examples. 
In \cref{sec:SingleParticle}, the simplest and most compact formulation of the optical device is presented. Single particle multiport interferometry is described by generalized Gell-Mann matrices, a connection to lattice path theory is drawn and approaching the problem by a recursive vector method, a solution algorithm is demonstrated that can easily calculate the necessary beam splitter and phase shifter values to have an MPI match any desired unitary operator. 
In \cref{sec:MultipleParticles}, the multi-particle case is presented for which a distinction into distinguishable and indistinguishable particles, as well as bosons and fermions is necessary. It is demonstrated that higher number states can be connected to higher dimensional matrix representations of the assosciated unitary Lie-groups.

%Recently, Clements~et~al.\cite{Clements16} redesigned the universal MPIs to reduce optical depth and increase robustness.

\section{Single Particle}
\label{sec:SingleParticle}

\subsection{The matrix approach}

Let's recapitulate the principle idea how to realize any discrete unitary operators in an interferometer. The arrangement of the MPI is given by a regular triangular mesh of two items: beam splitters and phase shifters.
In the mathematical description, two minor but essential deviations from the original of Reck~et~al.~\cite{Reck1994} are made here.

\begin{figure}
	\centering
	\iffalse
	\subfloat[][\centering3-port Mach-Zehnder type interferometer.]{\includegraphics[width=0.3\textwidth]{2port}\label{fig:AngleLattice0}}
	\hspace{7mm}
	\fi
	\subfloat[][\centering3-port Mach-Zehnder type interferometer.]{\includegraphics[width=0.4\textwidth]{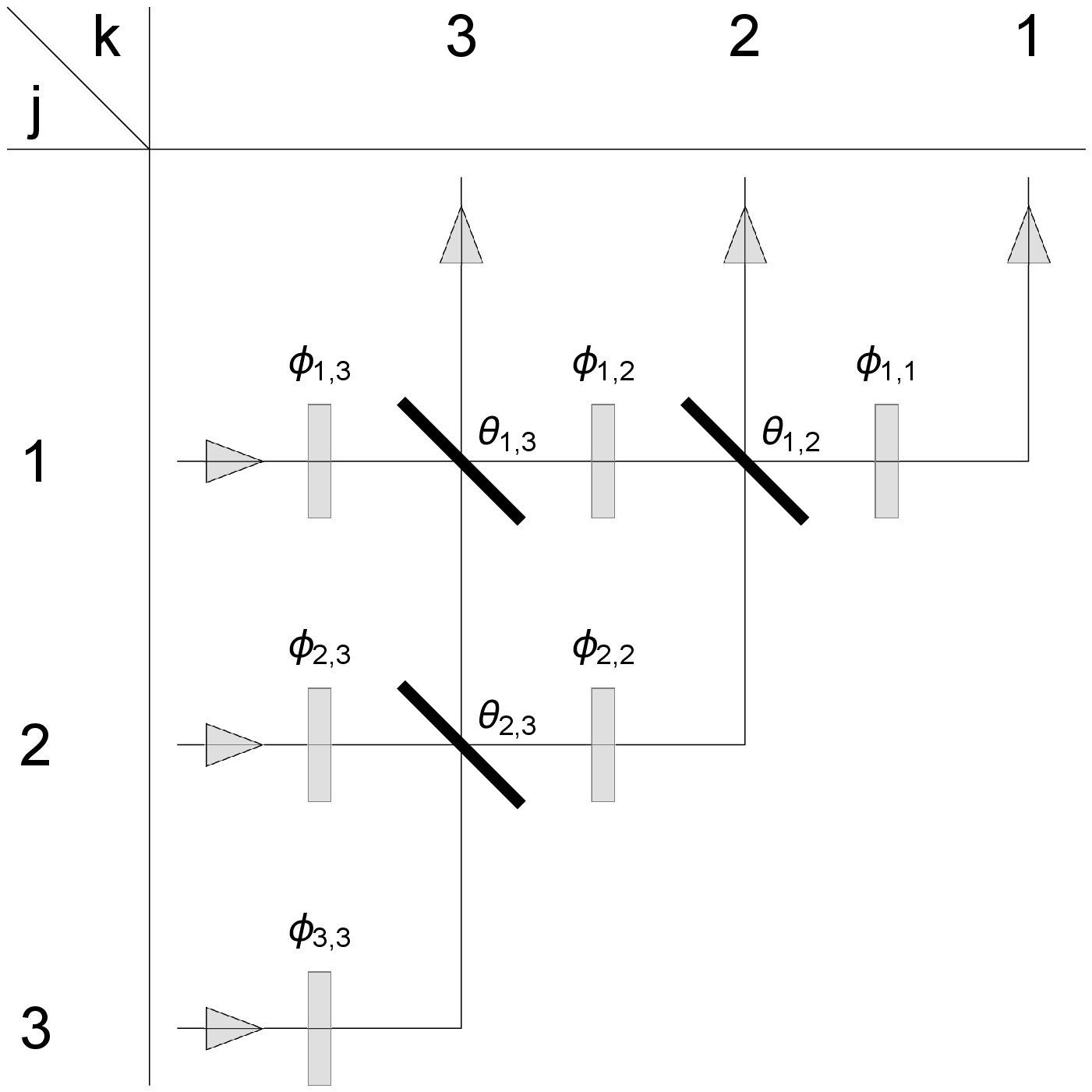}\label{fig:AngleLattice}}
	\hspace{7mm}
	\subfloat[][\centering5-port Mach-Zehnder type interferometer.]{\includegraphics[width=0.4\textwidth]{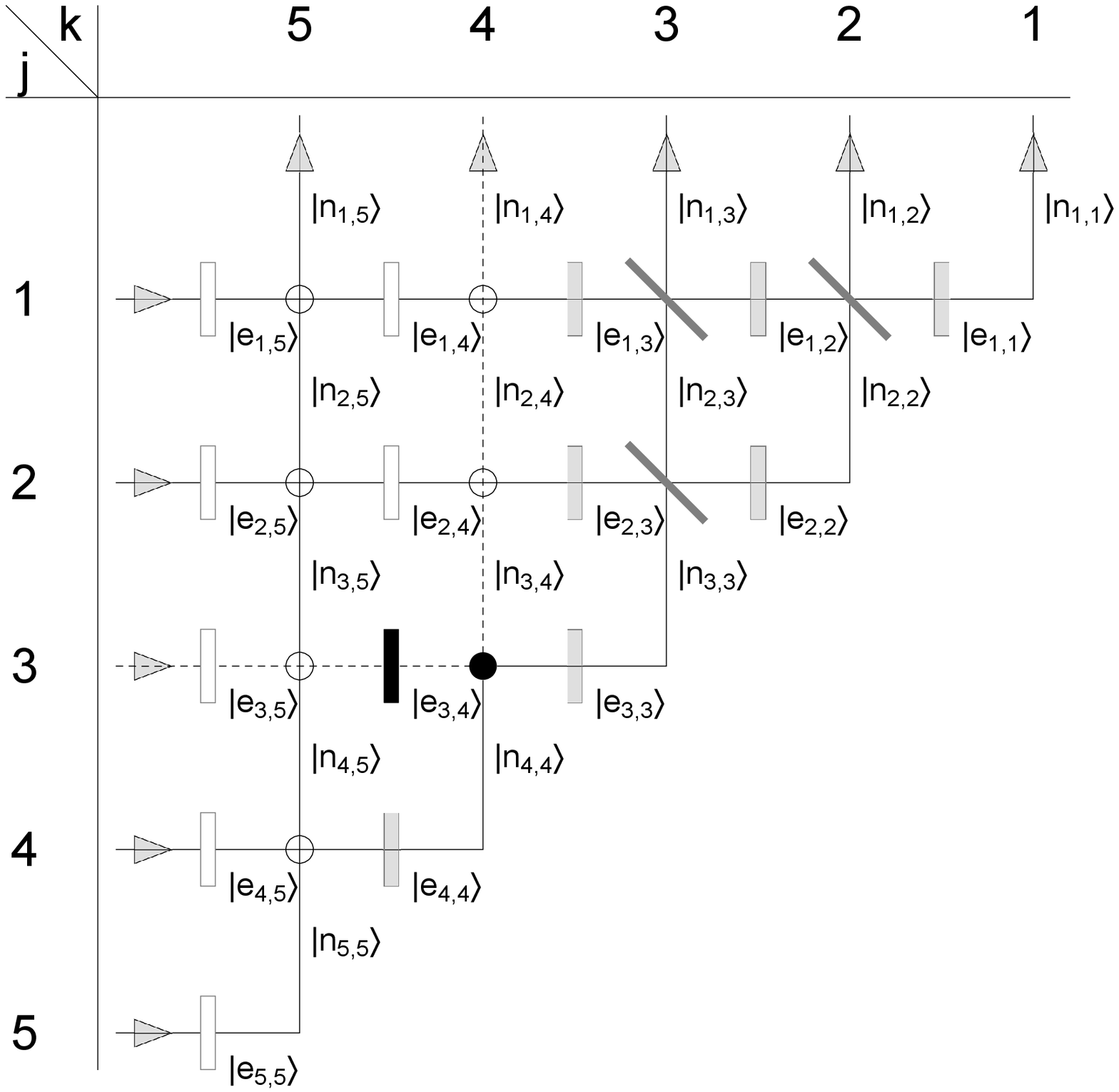}\label{fig:VectorLattice}}
	\caption{(a) On the left panel a multi-port interferometer, described by~\cref{eq:LatticeUnitary}, in dimension $d=3$ parametrized by the beam splitters~$\theta_{j,k}$ and phase shifters $\phi_{j,k}$ is shown. Note that the output ports (index~$k$) in the north-east lattice are labeled backwards. (b) The same expression for the unitary lattice operator can be obtained by consecutive superposition of the respective (complex) vectors according to~\cref{eq:StartCornerCondition,eq:Recursion}. In this $d=5$ dimensional example the black dot and rectangle indicate that the values $\theta_{3,4}, \phi_{3,4}$ are to be determined, in white what is calculated already and in gray the undetermined remaining part. The dashed path marks $D_{j,k}$ from~\cref{eq:Denom} excluding the black dot, everything within the dashed margin corresponds to $N_{j,k}$ from~\cref{eq:Nom}.}
	\label{fig:Pathlattice}
\end{figure}

\begin{enumerate}
	\item The lossless beam splitters are modeled by rotation matrices. For that purpose the anti-symmetric, generalized Gell-Mann matrices (GGM) (see e.g. \cite{Bertlmann2008}) are required
	
	\begin{equation}
	Y_{j,k}(n=1, d) \coloneqq -\frac{i}{2} \ket{j}\!\bra{k} + \frac{i}{2} \ket{k}\!\bra{j}~, \qquad 1\le j<k\le d~.
	\label{eq:GaGM}
	\end{equation}
	
	The first argument $n=1$ of the operator corresponds to the particle number and can be ignored until~\cref{sec:MultipleParticles}. In dimension two the matrix $Y_{1,2}(n=1, d=2)$ is equal to half the Pauli operator $\sigma_y$, in dimension~3 the three possible combinations give the following Gell-Mann matrices
	\begin{equation}
	Y_{1,2}(1,3) = \frac{\lambda_2}{2} = \frac{1}{2}\left(\begin{smallmatrix}	0 & -i & 0 \\ 	i & 0 & 0 \\ 	0 & 0 & 0	\end{smallmatrix}\right),\quad  Y_{1,3}(1,3) = \frac{\lambda_5}{2}= \frac{1}{2}\left(\begin{smallmatrix}	0 & 0 & -i \\ 	0 & 0 & 0 \\ 	i & 0 & 0	\end{smallmatrix}\right),\quad Y_{2,3}(1, 3) = \frac{\lambda_7}{2}  = \frac{1}{2}\left(\begin{smallmatrix}	0 & 0 & 0 \\ 	0 & 0 & -i \\ 	0 & i & 0	\end{smallmatrix}\right)~.
	\label{eq:Gellmann3}
	\end{equation}
	
	So, the anti-symmetric GGMs in~\cref{eq:GaGM} are a collection of $d \times d$ self-adjoint matrices with imaginary units at the positions indicated by the indices $(j,k)$. 	
	All beam splitters are henceforth described by the rotation matrices $\exp( i\,\theta_{j,k} Y_{j,k}(n,d))$ with angles~$\theta_{j,k}$. 
	
	The indices $(j,k)$ define the form~\cref{eq:GaGM} of the matrices in case of a single particle and likewise the position in the multi-port interferometer as shown in~\cref{fig:AngleLattice}. These unitary transformations describe a proper rotation since $\det(\exp( i\,\theta_{j,k} Y_{j,k}(n,d))) = +1$, while in~\cite{Reck1994} the determinant of the beam splitter matrices are chosen to be $-1$. Using the anti-symmetric over the symmetric GGM, $X_{i,j}(n=1,d) = \frac{1}{2}(\ket{j}\!\bra{k} + \ket{k}\!\bra{j})$, has the nice property that the entries of the exponential map $\exp( i\,\theta_{j,k} Y_{j,k}(n,d)) $ are all real (often it is preferred to give each reflected part a phase jump of $\frac{\pi}{2}$). 
	
	\item Interferometers contain phase shifters. It is counter-intuitive to place them at the output ports after the beam recombinations. So, as shown in~\cref{fig:Pathlattice} the phase shifters are placed in the horizontal beam paths.
	The phase shifters are likewise described by operators $\exp( i \,\phi_{j,k} E_{j}(n,d))$. The diagonal matrices $E_{j}(n=1,d) = \ket{j}\!\bra{j}$ tell in which `row' the phase shifts occur and do not require a separate `column' index.\\
\end{enumerate}

Let $A_i \in \mathbb{C}^{d \times d}$, then $\prod_{i}^n A_i = A_1 A_2 ...A_n$ is to be understood as the standard product of matrices in ascending order, where the components generally do not commute $[A_i,A_{i+1}] \neq 0$. The product notation allows to describe the entire multi-path interferometer for any particle number ($n=1$ in this section) and for arbitrary dimensions~\textit{d} of the interferometer in a very compact way as

\begin{equation}
U^{lat}(n, d) \coloneqq \prod_{\substack{k = d \\ k\texttt{--}}}^{1} e^{i \phi_{k,k} E_{k}(n,d)} \prod_{\substack{j = k-1 \\ j\texttt{--}}}^{1}e^{i \phi_{j,k} E_{j}(n,d)}  e^{i \theta_{j,k} Y_{j,k}(n,d) }~.
\label{eq:LatticeUnitary}
\end{equation}

This unitary operator $U^{lat}(d)$ describes the entire triangular-shaped interferometer.  The index of multiplication is given in descending order, indicated by the double minus in the index of multiplication, owing to the fact that the terms are written from left to right. Indeed,~\cref{eq:LatticeUnitary} is all it takes to write down the entire mixing of the discrete optical modes.\\

The fact that the product of these rotation matrices is a fitting model for the multi-port interferometer may seem surprising at first. For the correct interpretation it is good to write out~\cref{eq:LatticeUnitary} for dimension $d=3$,  abbreviate sine and cosine functions simply by $s(x)$ and $c(x)$ and set all $\phi_{j,k} = 0~\forall~\{j,k\}$ for the moment; the result is
\begin{equation}
U^{lat}(1, 3) = \left(
\begin{smallmatrix}
\text{c} \left(\frac{\theta_{1,2}}{2}\right) \text{c} \left(\frac{\theta _{1,3}}{2}\right) & \text{c} \left(\frac{\theta _{1,3}}{2}\right) \text{s} \left(\frac{\theta _{1,2}}{2}\right) & \text{s} \left(\frac{\theta _{1,3}}{2}\right) \\
-\text{c} \left(\frac{\theta _{2,3}}{2}\right) \text{s} \left(\frac{\theta _{1,2}}{2}\right)-\text{c} \left(\frac{\theta _{1,2}}{2}\right) \text{s} \left(\frac{\theta _{1,3}}{2}\right) \text{s} \left(\frac{\theta _{2,3}}{2}\right) & \text{c} \left(\frac{\theta _{1,2}}{2}\right) \text{c} \left(\frac{\theta _{2,3}}{2}\right)-\text{s} \left(\frac{\theta _{1,2}}{2}\right) \text{s} \left(\frac{\theta _{1,3}}{2}\right) \text{s} \left(\frac{\theta _{2,3}}{2}\right) & \text{c} \left(\frac{\theta _{1,3}}{2}\right) \text{s} \left(\frac{\theta _{2,3}}{2}\right) \\
\text{s} \left(\frac{\theta _{1,2}}{2}\right) \text{s} \left(\frac{\theta _{2,3}}{2}\right)-\text{c} \left(\frac{\theta _{1,2}}{2}\right) \text{c} \left(\frac{\theta _{2,3}}{2}\right) \text{s} \left(\frac{\theta _{1,3}}{2}\right) & -\text{c} \left(\frac{\theta _{2,3}}{2}\right) \text{s} \left(\frac{\theta _{1,2}}{2}\right) \text{s} \left(\frac{\theta _{1,3}}{2}\right)-\text{c} \left(\frac{\theta _{1,2}}{2}\right) \text{s} \left(\frac{\theta _{2,3}}{2}\right) & \text{c} \left(\frac{\theta _{1,3}}{2}\right) \text{c} \left(\frac{\theta _{2,3}}{2}\right)
\end{smallmatrix}
\right).
\label{eq:Ulatd3}
\end{equation}

In~\cref{eq:Ulatd3} read multiplication as logical AND, addition as OR, then $\sin\frac{\theta_{i,j}}{2}$ as left-turn, $-\sin\frac{\theta_{i,j}}{2}$ as right-turn and $\cos\frac{\theta_{i,j}}{2}$ as going straight, \cref{eq:LatticeUnitary} is then nothing but a symbolic way of expressing all possible combinations of North-East lattice paths where rows correspond to the $j$-th input port and columns to the $k$-th output labeled in reverse order as shown in~\cref{fig:Pathlattice}. For example, the entry in the second row, first column provides the information how to get from the 2nd input to the output labeled as `1'.\\ 

A few remarks: first, any interferometer that can be embedded into the triangular lattice can also be described by~\cref{eq:LatticeUnitary} when the shape is deliberately changed by `erasing a node' by setting $\theta_{j,k} = 0$ or $\pi/2$  (node will be used synonym to beam splitter from now on). More generally, the form of the interferometer is not relevant for the validity of the description, \cref{eq:LatticeUnitary} is true as long as the paths are homotopic relative to the beam splitters. It is only when we want to have a real physical implementation, that we need a rigid geometric configuration. Second, instead of a `transmissive labeling' it is possible to use a `reflective labeling' where the paths keeps their numbering along the reflected route. In that case~\cref{eq:LatticeUnitary} takes on a form that is more reminiscent of an Euler parametrization for the unitary groups~\cite{Tilma2002}.
Finally, the number of Dyck paths of order~$m$ in a square lattice is given by the Catalan numbers $C_m = \frac{1}{m+1} \binom{2m}{m}$. The generalization of $C_m$, called Catalan's trapezoid numbers~\cite{shlomi2014}
\iffalse
\begin{equation}
C_{\alpha} ({\beta},{\gamma}) = \left\{\begin{array}{cl} \binom{{\beta}+{\gamma}}{{\gamma}} \qquad  & 0 \le {\gamma} < {\alpha} \\\binom{{\beta}+{\gamma}}{{\gamma}}-\binom{{\beta}+{\gamma}}{{\gamma}-{\alpha}} \qquad & {\alpha} \le {\gamma} \le {\beta}+{\alpha}+1 \\ 0 \qquad & {\gamma} > {\beta}+{\alpha}+1  \end{array} \right.
\label{eq:CatalansTrapezoid}
\end{equation}\fi
yields the correct number of paths in the MPI %The number of summands of the respective matrix elements $\braket{j|U^{lat}(d)|k}$ is $C_{(d-j+1)} (j-1,d-k)$, $1 \le j \le d$, $1 \le k \le d$, 
and gives a good idea as to how quickly the complexity and the number of terms in the unitary lattice operator increase.

\subsection{The recursive appraoch}

The true power of the quantum MPI is that it allows to realize any finite unitary operator. The aim must therefore be the calculation of all phase shifts $\phi_{j,k}$ and beam splits $\theta_{j,k}$ to have~\cref{eq:LatticeUnitary} match any desired unitary, target matrix $U^{lat}(n=1, d) = U^{tar}$. In the following lines, a computationally feasible algorithm is presented that is completely equivalent to the previous matrix method.\\

The highly regular and ordered structure of the lattice allows the solution to be represented symbolically by a vectorial recursion. For that purpose the east and north going paths are all indicated by kets $\ket{e_{j,k}}$ and $\ket{n_{j,k}}$ as illustrated in~\cref{fig:VectorLattice}. Let's first establish the starting condition and notice that the vectors at the lattice corners are identical up to a phase factor

\begin{equation}
\ket{e_{j,d}} \coloneqq \ket{j}~, \qquad \ket{n_{j,j}} \coloneqq e^{i \phi_{j,j}} \ket{e_{j,j}}  \qquad 1 \leq j \leq d~.
\label{eq:StartCornerCondition}
\end{equation}

Subsequently, all paths intersect according to the rule
\begin{equation}
\left.
\begin{array}{r}
\ket{n_{j,k}} \coloneqq e^{i \phi_{j,k}}\sin \left(\frac{\theta_{j,k}}{2}\right)\ket{e_{j,k}} + \cos \left(\frac{\theta_{j,k}}{2}\right)\ket{n_{j+1,k}} \\
\ket{e_{j,k-1}} \coloneqq e^{i \phi_{j,k}}\cos \left(\frac{\theta_{j,k}}{2}\right)\ket{e_{j,k}} - \sin \left(\frac{\theta_{j,k}}{2}\right)\ket{n_{j+1,k}} 
\end{array} \quad \right\}\\ \quad 1\le j \le k\le d~.
\label{eq:Recursion}
\end{equation}

As it turns out, the set of all vectors at the output port yield exactly the unitary matrix~\cref{eq:LatticeUnitary}
\begin{equation}
U^{lat}(1,d) = \left\{\ket{n_{1,k}}^T\right\}_{k=1}^d = \left\{\ket{n_{1,1}},\ket{n_{1,2}},\ldots,\ket{n_{1,d}}\right\}^T~.
\label{eq:UnitaryVectors}
\end{equation}

All that is left to do now is to rearrange the formula to have the desired term on one side.  At this point it is also good to use Euler's formula $e^{i \phi_{j,k}} = (x_{j,k}+i y_{j,k})$ for all phases and make the substitutions $\sin \left(\theta_{j,k}/2\right) = \sqrt{R_{j,k}} $ and $\cos \left(\theta _{j,k}/2\right) = \sqrt{1-R_{j,k}} $ as the products of square roots are easier to handle than trigonometric identities. Writing out a few terms of the iteration reveals a pattern from which the general solution can be obtained

\begin{equation}\label{eq:Solution}
\begin{aligned}
R_{j,k} = \left|Z_{j,k}\right|^2~,  \qquad x_{j,k} &= \frac{\text{Re} (Z_{j,k})}{\sqrt{R_{j,k}}}~, \quad y_{j,k} = \frac{\text{Im} (Z_{j,k})}{\sqrt{R_{j,k}}}~, \quad 1\le j<k \le d~. \\
x_{kk} &= \text{Re} (Z_{kk})~, \hspace{5mm} y_{kk} = \text{Im} (Z_{kk})~.
\end{aligned}
\end{equation}

where the reflectivities $R_{j,k}$ at the respective nodes are given by the modulus of the quotient

\begin{equation}
Z_{j,k} \coloneqq \frac{U^{tar}_{j,k} - N_{j,k}}{D_{j,k}}  
\label{eq:DefintionOfZ}
\end{equation} 

which requires the defintions of the numerator and denominator term
\begin{subequations}
\begin{align}
N_{j,k} &\coloneqq \sum\limits_{m=1}^{j-1}e^{i \phi_{m,k}}\sqrt{R_{m,k}}\braket{j|e_{m,k}}\prod\limits_{l=1}^{m-1}\sqrt{1-R_{l,k}}~, \label{eq:Nom} \\
D_{j,k} &\coloneqq \braket{j|e_{j,k}}\prod\limits_{l=1}^{j-1}\sqrt{1-R_{l,k}}~. \label{eq:Denom}
\end{align}
\end{subequations}

\cref{eq:Solution} can be used to determine all values of the entire MPI in an ordered way, proceeding in the form $(j,k) = \{(1,d)\rightarrow (2,d)\rightarrow\dots\rightarrow(d-1,d)\rightarrow(1,d-1)\rightarrow(2,d-1)\dots \}$. It is sufficient to solve the $\binom{d}{2}$ entries above the diagonal recursively from top to bottom and then from right to left, provided that the parameters (e.g. $\theta_{1,3} = \pi$) do not vanish. It is crucial to keep the non-bijective nature of the functions involved in mind, which makes a sign convention very important when determining the angles as otherwise errors occur in the arithmetic continuation of the computation.\\

For an intuitive understanding of the solution its good to look at~\cref{fig:VectorLattice}. 
In the example the values to be calculated $(\theta_{3,4}, \phi_{3,4})$ are the black dot on one corner and the preceding phase shifter. Since the algorithm works according to a strict sequence, this implies that all beam splitters to the left and above this node are already determined, indicated by white fillings. The dashed path marks $D_{j,k}$. The expression $\braket{j|e_{m,k}}$ is to be interpreted as the set of all paths from input port $j$ to the point $(m, k)$ excluding this node (in~\cref{fig:VectorLattice} (m,k) = (3,4))). The remainder $\prod_{l=1}^{m-1}\sqrt{1-R_{l,k}}$ corresponds to the remaining northwards dashed path. The expression $N_{j,k}$ in turn characterizes all routes contained in the dashed margin. \\

The solvability of the linear optical MPI depends on the design of the lattice.
The set of paths from an input to an output port corresponds to one equation. In the triangular form it is apparent that it is possible to go from input port $j$ to output port $k$ such that it contains only one unknown $\phi_{j,k}$ at the edge leading right into one unknown $\theta_{j,k}$ sitting at the bottom right node. Mathematically, this means every splitter and phase value is determined one at a time by a single equation. 
The presented algorithm fails in two extreme cases, either when $R_{j,k}= 0$ or $R_{j,k}= 1 \rightarrow D_{j,k}= 0$. A workaround for these situations is presented in the~\cref{app:Appendix1}.

\subsubsection{Examples}

\begin{table}
	\begin{adjustbox}{width=\textwidth}
		\begin{tabular}{cl}\hline
			Reflec.& \begin{tabular}{llllll}
				$R_{1,7} \approx 0.143$ & $R_{2,7} \approx 0.167$ & $R_{3,7} = 0.2$ & $R_{4,7}=0.25$ & $R_{5,7}\approx0.333$ & $R_{6,7}= 0.5$ \\ $R_{1,6} = R_{2,7}$ & $R_{2,6} \approx 0.254$ & $R_{3,6} \approx 0.371$ & $R_{4,6} \approx 0.519$ & $R_{5,6} \approx 0.714$ &  \\ $R_{1,5} = R_{3,7}$ & $R_{2,5} = R_{3,6}$ & $R_{3,5}\approx 0.565$ & $R_{4,5}\approx0.771$ & \\
				$R_{1,4} = R_{4,7}$ & $R_{2,4} = R_{4,6}$ & $R_{3,4}= R_{4,5}$ \\ $R_{1,3} = R_{5,7}$ & $R_{2,3} = R_{5,6}$ & \\ $R_{1,2} = R_{6,7}$ &
			\end{tabular}\\\hline 
			phases & \begin{tabular}{llllll}
				$\phi_{2,7}\approx 5.386$ & $\phi_{3,7}\approx 4.488$ & $\phi_{4,7}\approx 3.590$ & $\phi_{5,7}\approx 2.693$ & $\phi_{6,7}\approx 1.795$ & $\phi_{7,7}\approx 0.898,$\\ $\phi_{2,6}\approx 5.503$ & $\phi_{3,6}\approx 4.802$ & $\phi_{4,6}\approx 4.150$ & $\phi_{5,6}\approx 3.525$ & $\phi_{6,6}\approx 2.917$ & \\ $\phi_{2,5}\approx 5.582,$ &$\phi_{3,5}\approx 4.939$ & $\phi_{4,5}\approx 4.300$ & $\phi_{5,5}\approx 3.656$ \\  $\phi_{2,4}\approx 5.630$ & $\phi_{3,4}\approx 4.992$ & $\phi_{4,4}\approx 4.341,$\\$\phi_{2,3}\approx 5.659$ & $\phi_{3,3}\approx 5.014$ \\ $\phi_{2,2}\approx 5.675$
			\end{tabular}\\\hline 
		\end{tabular}
	\end{adjustbox}
	\caption{Numeric results for a multi-port interferometer realizing a 7 dimensional Fourier transform. The upper row shows the probabilities for reflection $R_{j,k} = \sin^2(\theta_{j,k}/2)$ where the first probabilities satisfy $R_{k,7} = 1/(8-k)$. The results are symmetric about the diagonal in the MPI. In the lower row the phase shifter values $\phi_{j,k}$ are given. Since the first row in the Fourier transform is real, $\phi_{1,k} = 0~\forall k$.}
	\label{tab:FourierTable}
\end{table}

\textbf{Fourier transformation:} As a numerical example the targeted unitary transformation is a discrete Fourier transform given by

\begin{equation}
U^{tar} = F(d) \coloneqq  \frac{1}{\sqrt{d}}\sum_{\{j,k\}=1}^{d} e^{\frac{2\pi i}{d}(j-1)(k-1)} \ket{j}\!\bra{k}~.
\end{equation}

%In quantum physics, the discrete Fourier transformation is an integral part of information theoretic algorithms. 
\cref{tab:FourierTable} shows the reflection probabilities and phase shifts necessary to realize the Fourier transform in a multiport interferometer of dimension~7. \\

\textbf{Wigner d-transformation:}
As an analytic example, consider the algebra $\mathfrak{su}(2)$ with its three generators $S_x, S_y, S_z$ obeying the commutation relation $\left[S_x, S_y\right] = i \epsilon^{xyz} S_z \,$ with structure constants equal to the Levi-Civita symbol. Choosing $S_z$ to be the diagonal element it's eigenvalue equation is given by $S_z \ket{s, m_s}_z = m_s \ket{s, m_s}_z \,$
The finite dimensional, irreducible representation (irrep) of the algebra are labeled by the  integer and half-integer valued quantum numbers $s$ and $m_s = \{-s,-s+1,...,s+1,s\}$.\\

For a nice and useful analytic example the next target unitary operators are the Wigner (small) d-matrices, the $d = (2s+1)$-dimensional unitary representation of SU(2) that can be written succinctly as
\begin{equation}
U^{tar} = d_s(\theta) \coloneqq \sum_{\{j,k\}=1}^{2s+1}\braket{j\vert e^{i \theta S_y}\vert k}\ket{j}\!\bra{k}~.
\end{equation} 

Since the d-matrices are all real valued, there is no need to consider phase shifts; also for a more compact notation of the solutions it is useful to make the substitution $\sin^2\frac{\theta}{2} \mapsto t$. \cref{tab:Wigner1DTable} lists the reflectivities $R_{i,j}(t)$ up to dimension 5. %Physically, this example is interesting in its own since the results also make up the building blocks for other group transformations. 
One should be aware that in such a setup all beam splitters are coupled to each other and cannot be tuned independently. The solutions are acquired by the algorithm~\cref{eq:Solution}, however from the look of it a general analytic solution seems plausible.

\begin{table}
	\begin{adjustbox}{width=\textwidth}
	\begin{tabular}{l|llllllll}
		{Dim} & \multicolumn{6}{c}{Reflectivities} \\\hline
		2 &$R_{1,2} = t$ & \\
		3 &$R_{1,3} = t^2$ & $R_{2,3} = \frac{2 t}{t+1}$\\
		4 &$R_{1,4} = t^3$ & $R_{2,4} = \frac{3 t^2}{t^2+t+1}$ & $R_{3,4} = \frac{3 t}{2 t+1}$ & $R_{2,3} = \frac{t (t+2)^2}{(2 t+1)^2}$ \\
		5 &$R_{1,5} = t^4$ & $R_{2,5} = \frac{4 t^3}{t^3+t^2+t+1}$ & $R_{3,5} = \frac{6 t^2}{3 t^2+2 t+1}$ & $R_{4,5} = \frac{4 t}{3 t+1}$ & $R_{2,4} = \frac{t^2 \left(t^2+2 t+3\right)^2}{\left(3 t^2+2 t+1\right)^2}$ & $R_{3,4}=\frac{6 t (t+1)^2}{(3 t+1) \left(t^2+4 t+1\right)}$
%		6 &$R_{1,6} = t^5$ & $R_{2,6} = \frac{5 t^4}{t^4+t^3+t^2+t+1}$ & $R_{3,6} = \frac{10 t^3}{4 t^3+3 t^2+2 t+1}$ & $R_{4,6} = \frac{10 t^2}{6 t^2+3 t+1}$ & $R_{5,6} = \frac{5 t}{4 t+1}$ $R_{2,5}=\frac{t^3 \left(t^3+2 t^2+3 t+4\right)^2}{\left(4 t^3+3 t^2+2 t+1\right)^2}$ & $R_{3,5}=\frac{2 t^2 \left(3 t^2+4 t+3\right)^2}{\left(6 t^2+3 t+1\right) \left(t^4+4 t^3+10 t^2+4 t+1\right)}$ & $R_{4,5}=\frac{2 t (3 t+2)^2}{(4 t+1) \left(3 t^2+6 t+1\right)}$ & $R_{3,4}=\frac{t \left(t^2+6 t+3\right)^2}{\left(3 t^2+6 t+1\right)^2}$
	\end{tabular}
	\end{adjustbox}
	\caption{Analytic results for the probabilities of reflection for the target unitary given by the single parameter Wigner rotation matrix in dimension $2s+1$. For a clearer view, the d-matrix elements are substituted by $\sin\frac{\theta}{2} \mapsto \sqrt{t}$ and $\cos\frac{\theta}{2} \mapsto \sqrt{1-t}$ with $t \in [0,1]$ . Like~\cref{tab:FourierTable}, this table also makes use of the symmetry $R_{j,k}= R_{d+1-k,d+1-j}$ and lists the results in the order the values are calculated.}
	\label{tab:Wigner1DTable}
\end{table}

\section{Multiple Particles}
\label{sec:MultipleParticles}

The previous section of the manuscript was dedicated to single particles and as such holds for all idealized quantum systems in principle. But as soon as there is a second particle, it is necessary to make the distinction between fermions (\textbf{F}) and bosons (\textbf{B}), differentiate between particles that are distinguishable (\textbf{D}) or not. As already indicated, \cref{eq:LatticeUnitary} is in its final form, it is only necessary the replace the GGMs and phase shifter matrices that describe $n=1$ particle with the correct matrices $Y_{j,k}(n,d)$ and $E_{k}(n,d)$ for a proper description of multi-particle MPIs.

\subsection{Indistinguishable bosons}

In the following lines bosons, characterized as number (Fock) states with no distinguished properties are considered. 
As a first step a representation in terms of a rotation matrix in the case of a single beam splitter for two bosons (2B) is sought. Some discrete hints for the proper description were indicated already. For a better understanding a closer look at the Wigner-d transformation is imperative. \\

Let the $\mathfrak{su}(2)$ matrices be written as $S_i(2s+1)$, then for $s = \frac{1}{2}$ the operator $Y(n=1, d=2) = S_y(2) =\sigma_y/2$ yields half the $y$-Pauli matrix that is used for a single particle, where the node indices have been omitted for the moment.  The exponential map of the $s=1$ matrix

\begin{equation}
\begin{gathered}
\hspace{2.5cm}\begin{array}{ccc}\ket{2,0} \hspace{1.5cm} & \ket{1,1} \hspace{1.5cm}&\ket{0,2}\end{array}\\ 
e^{i \theta Y(2B,2)} = e^{i \theta S_y(3)} = \left(
\begin{array}{ccc}
\cos ^2\frac{\theta }{2} & \sqrt{2}\cos \frac{\theta }{2} \sin \frac{\theta }{2} & \sin ^2\frac{\theta }{2} \\
-\sqrt{2}\cos \frac{\theta }{2} \sin \frac{\theta }{2} & \cos^2 \frac{\theta }{2} - \sin^2 \frac{\theta }{2} & \sqrt{2}\cos \frac{\theta }{2} \sin \frac{\theta }{2} \\
\sin ^2\frac{\theta }{2} & -\sqrt{2}\cos \frac{\theta }{2} \sin \frac{\theta }{2} & \cos ^2\frac{\theta }{2} \\
\end{array}
\right) \quad\begin{array}{c}
\ket{2,0} \\ \ket{1,1} \\ \ket{0,2}
\end{array}
\end{gathered}
\label{eq:2Boson1BS}
\end{equation}

gives exactly the expected form for two bosons coincident on a single beam splitter. The number states at the top and right edges highlight how the input and output modes are connected.  Conveniently, the previous interpretation as symbolic path information sill holds: $\cos ^2\frac{\theta }{2}$ and $\sin ^2\frac{\theta }{2}$ correspond to both particles  being jointly either transmitted or reflected. The term $\sqrt{2}\cos \frac{\theta }{2} \sin \frac{\theta }{2}$ indicates that one boson goes straight while the second one turns left or vice versa. The coefficient $\sqrt{2}$ is characteristic for the $\{2\} \leftrightarrow \{1,1\}$ mode mixing and can be calculated by the explicit form of the Wigner d-matrix~\cite{Morrison1987}, or alternatively from (appendix in~\cite{Ou2007})

\begin{equation}
	\ket{\psi(M,N)}_{out} = \frac{1}{\sqrt{M! N!}} (\cos\frac{\theta }{2} a_1^{\dagger}-\sin\frac{\theta }{2} a_2^{\dagger})^M (\cos\frac{\theta }{2} a_2^{\dagger}+\sin\frac{\theta }{2} a_1^{\dagger})^N \ket{0,0}~,
\end{equation}

where $a_1^{\dagger}, a_2^{\dagger}$ are the creation operators of power $M$ and $N$.
Finally in the middle the mode $\ket{1,1} \rightarrow \ket{1,1}$ is obtained when either both bosons go straight or when one makes a right, the other particle a left turn at the beam splitter. Utilizing the trigonometric identity $\cos^2 \frac{\theta }{2} - \sin^2\frac{\theta }{2} = \cos \theta$ apparently the famous Hong-Ou-Mandel (HOM) effect is obtained when $\theta$ is an integer multiple of $\frac{\pi}{2}$ (respectively $R=\sin^2 \frac{\pi}{4}$).\\

Thus, any number of bosons that are coincident on a single beam splitter can be described by a compact expression as short as $e^{i \theta Y(n = 2s,2)} = e^{i \theta S_y(2s+1)}$, where the number of bosons $n = 2s$ changes the dimension $2s+1$ of the algebraic representation. In retrospective, \cref{tab:Wigner1DTable} highlights the equivalency of $n$-boson two mode interference, to a single-particle $d+1$ mode interferometer of the triangular form. If for example the three beam splitters in the set-up~\cref{fig:AngleLattice} are set to $R_{1,3}=\sin^4\frac{\theta}{2}$ and $R_{2,3} = R_{1,2} = 2 \sin^2\frac{\theta}{2}/(\sin^2\frac{\theta}{2}+1)$ then a two boson single beam splitter interference is simulated. The one-to-one correspondence between the states \{$\ket{2,0} \leftrightarrow \ket{1,0,0}$, $\ket{1,1} \leftrightarrow \ket{0,1,0}$, $\ket{0,2} \leftrightarrow \ket{0,0,1}$\} therefore implies that  for $\theta = \frac{\pi}{2}$ ($R_{1,3}= 1/4 , R_{2,3} = R_{1,2} = 2/3$) a HOM effect is imitated by complete destructive interference of the event $\ket{0,1,0}$. \\

As SU(2) is used for a single beam splitter and the higher dimensional irreps correspond to more particles, intuition suggests a logical continuation to larger unitary groups SU(d). To find explicit matrix representation in the desired `lattice basis', a construction scheme is presented in the following subsection.\\

\subsubsection{Construction scheme}
\label{sec:ConstructionScheme}

The general approach to create the required matrices is illustrated along the example of a 2-boson 3-port set-up. \\

1. For the start the set of all possible Fock states are collected in an ordered set

\begin{equation}
\ket{\text{co}(nB,d)}_i \coloneqq \left\{\ket{n_1, n_2 , \ldots , n_d} \vert \text{canoncially orderd},  \sum_{a=1}^{d} n_a = n, n_a \in \mathbb{N}_0 \right\}_{i \in I}
\label{eq:CoBoson}
\end{equation}

So for the example $(n,d) = (2B,3)$,
\begin{equation}
\begin{split}
\ket{\text{co}(2B,3)} &= \left\{\ket{2,0,0}, \ket{1,1,0}, \ket{1,0,1}, \ket{0,2,0}, \ket{0,1,1}, \ket{0,0,2} \right\}~.%  = \{\vec{e}_1, \vec{e}_2, \vec{e}_3, \vec{e}_4, \vec{e}_5, \vec{e}_6 \}
\end{split}
\end{equation}
The number of states with $n$-bosons in $d$-path modes in general is given by the Binomial coefficient  $\binom{d-1+n}{n}$. The ordering scheme keeps the numbers in the first slot larger than the number in the next slot and keeps permuting by that rule. The ordering is correct if by applying the particle number operator $N_d = a^{\dagger}_d a_{d}$ to the last element of each Fock state the set $\bra{\text{co}(2B,3)}  N_{d=3} \ket{\text{co}(2B,3)}  = \{0, 0,1, 0,1,2\}$ allows for a partition $P$ into blocks of integers strictly increasing by one $P(\{0, 0,1, 0,1,2\}) \rightarrow P_1 = \{0\}, P_2 = \{0,1\}, P_3 = \{0,1,2\}$.\\

2. In the next step a swap operation $SW_{j,k}$ is applied on the set of ordered number states. 
This simple procedure changes the position of the slots in the Fock basis $n_j \rightleftarrows n_k$, e.g.
\begin{equation}
\begin{split}
SW_{1,2}\ket{\text{co}(2B,3)} &= \left\{SW_{1,2}\ket{2,0,0}, SW_{1,2}\ket{1,1,0}, SW_{1,2}\ket{1,0,1}, SW_{1,2}\ket{0,2,0}, SW_{1,2}\ket{0,1,1}, SW_{1,2}\ket{0,0,2} \right\} \\ &= \{\ket{0,2,0}, \ket{1,1,0}, \ket{0,1,1}, \ket{2,0,0}, \ket{1,0,1},\ket{0,0,2} \}~. %=  \{\vec{e}_4, \vec{e}_2, \vec{e}_5, \vec{e}_1, \vec{e}_3, \vec{e}_6 \}
\end{split}
\end{equation}
With that the permutation matrices $\Pi_{ij}$ are calculated
\begin{equation}
\begin{split}
\Pi_{j,k} = \Pi_{k,j} \coloneqq \sum_{i=1}^{\binom{d-1+n}{n}}SW_{j,k}\ket{\text{co}(n,d)}_i \tensor[_i]{\bra{\text{co}(n,d)}}{}~, \qquad 1\le j<k\le d~.
\end{split}
\label{eq:Permutes}
\end{equation}

In this example this gives three permutation matrices $\Pi_{1,2}, \Pi_{1,3}$ and $\Pi_{2,3}$.\\

3. Finally, the previous two steps are combined to construct the required $\binom{d}{2}$ operators.
The first element is obtained by constituting the block diagonal matrix

\begin{gather}
Y_{d-1,d} \coloneqq  \bigoplus_{i=1} S_y(\vert P_i \vert) \rightarrow  Y_{2,3}(2B,3) = S_y(1) \oplus S_y(2) \oplus S_y(3)~,
\label{eq:LastBS}
\end{gather}

where the direct sum is taken over the number of partitions and the dimension of the representation is given by the partitions' cardinalities. All other elements are related to the first one $Y_{d-1,d}$ or succeeding elements by permutation
\begin{equation}
Y_{j,k} = \Pi_{i,j} Y_{i,k} \Pi_{i,j} = \Pi_{i,k} Y_{j,i} \Pi_{i,k}~,
\label{eq:RestBS}
\end{equation}

whereby all beam splitter operators are obtained. Because swapping the indices in~\cref{eq:RestBS} makes things easier, it is necessary to formally define 
\begin{equation}
A_{k,j}(n,d) \coloneqq A^{T}_{j,k}(n,d)
\end{equation}

where $A$ can be $X$ or $Y$. This is natural for one particle as in the case for the GGMs~\cref{eq:GaGM}, however for $n>1$ these `lattice subscripts' $(j,k)$ no longer correspond to matrix indices.
Accordingly, $Y_{i,k} = - Y_{k,i}$ are anti-symmetric, the X's of course $X_{j,k} = X_{k,j}$ are not. The phase shifters are far simpler to determine and are given by the $d$-diagonal matrices

\begin{equation}
E_k(n,d) \coloneqq \sum_{i=1}^{\binom{d-1+n}{n}} \tensor[_i]{\bra{\text{co}(nB,d)}}{} N_k \ket{\text{co}(nB,d)}_i \ket{i}\bra{i}~.
\label{eq:AllPS}
\end{equation}

The results of this particular example can be found in the~\cref{app:Appendix2}.
Thus, the entire generation of the operators is completed and the interferometer with any number of paths~$d$ and particles~$n$ is obtained by insertion of~\cref{eq:LastBS,eq:RestBS,eq:AllPS} into~\cref{eq:LatticeUnitary}.
\begin{center}
	$\ast$~$\ast$~$\ast$
\end{center}
To see whether the initial intuition about particle numbers and the dimension of the irreps is true the following must be done; first label $Y_{1,2}(2B,3) =T_2$,  $Y_{1,3}(2B,3) =T_5$ and  $Y_{2,3}(2B,3) =T_7$ by a single algebraic index, then repeat the procedure with the symmetric elements $X_{1,2}(2B,3) =T_1$, where in~\cref{eq:LastBS} $S_y$ is replaced by $S_x$ (or by taking the modulus of the $Y_{i,j}$ matrix entries), and of course $X_{1,3}(2B,3) =T_4$, $X_{2,3}(2B,3) =T_6$. Finally two diagonal operators $T_3 = -i[T_1,T_2]$ and $T_8 = -(T_3 + 2i[T_4,T_5] )/\sqrt{3}$ can be generated. This results in 8 6x6 matrices (see~\cref{app:Appendix2}) whose commutator relation $[T_a,T_b] = i f^{abc} T_c$ have the structure constants of $\mathfrak{su}(3)$. In conclusion, the presented scheme, as anticipated, leads to the construction of the off-diagonal parts of the respective Lie algebras where the number of ports $n$ corresponds to the dimension in the algebraic series, while the number of particles is related to the dimension of the matrix representations.\\

For $\mathfrak{su}(3)$ the irreps have dimension $D(p,q) = \frac{1}{2}(p+1)(q+1)(p+q+2)$ \cite{Hall2015} which has to match the formula $\binom{3-1+n}{n} = (n+1)(n+2)/2$. The equivalency is given when $q=0$ (or $p=0$) which implies that there are missing representations (e.g. the adjoint representation $p=q=1$), a fact that pertains to the higher dimensional algebras. In this context, it seems that only the one-dimensional subspaces of the representations have an appropriate meaning. This is only logical, since the dimension of the representation is determined by a single parameter, the particle number $n$.

\subsection{Indistinguishable fermions}

It is now possible to switch to the case of fermions with minimal effort by modifying the construction scheme in~\cref{sec:ConstructionScheme}. All that is necessary is to change the ruling for the possible Fock states from~\cref{eq:CoBoson} to

\begin{equation}
\ket{\text{co}(nF,d)}_i \coloneqq \left\{\ket{n_1, n_2 , \ldots , n_d} \vert \text{canoncially orderd},  \sum_{a=1}^{d} n_a = n, n_a \in \{0,1\} \right\}_{i \in I}
\label{eq:CoFermion}
\end{equation}

The limitation of $n_a$ to $\{0,1\}$ naturally means that the number of fermions cannot exceed the number of paths $n\leq d$. One substantial consequence of this limitation is that the MPIs' number of anti-symmetric matrices and their respective sizes are given by the Binomial coefficient $\binom{d}{n}$ which naturally also changes the upper limit of summation in~\cref{eq:Permutes,eq:AllPS}.
With the construction~\cref{eq:CoFermion} the probability amplitude of the $\ket{1,1}$ mode is fixed 
\begin{equation}
p(\ket{1,1} \rightarrow \ket{1,1}) = 1 = |e^{i \theta Y(2F,2)}|^2 \xRightarrow[]{\text{fixed}}  Y(2F,2) = 0~.
\label{eq:Fermion11prob}
\end{equation}

%With our interpretation~\cref{eq:Fermion11prob} suggests $\ket{1,1}$ is always transmitted and reflection is prohibited. \\

Now let's see for examples. For two indistinguishable fermions in three ports, we have $\binom{3}{2} = 3$, therefore three 3x3 matrices are all that is necessary to mathematically describe the beam splitters in the interferometer. A quick calculation yields $Y_{1,2}(2F,3) = \lambda_7/2$, $Y_{1,3}(2F,3)  = \lambda_5/2$ and $Y_{2,3}(2F,3)  = \lambda_2/2$. Once again the anti-symmetric Gell-Mann matrices have been received, however the order is reversed with respect to~\cref{eq:Gellmann3}. The reason for that is a change in the meaning of the zero in the matrices, i.e looking  at $\lambda_2$ for example the last '$0$' in the diagonal is no longer to be interpreted as no particle at the beam splitter, but instead as two coincident fermions in conformity with~\cref{eq:Fermion11prob}. The phase shifts can be obtained by the same concept as~\cref{eq:AllPS}.
In the case of two identical fermions in a 4-port interferometer there are $\binom{4}{2}$ anti-symmetric matrices as shown in the~\cref{app:Appendix2}. \\

A serious problem that appears with these and higher dimensional fermionic examples is that there is no closure under commutation, the reason being  unanticipated sign changes in the commutator. By introducing

\begin{equation}
\eta_{1,2}(nF,d)  \coloneqq \left\{
\begin{array}{lr}
\mathbbm{1}_{\binom{d}{n}}~,  &n=1 ~,\\
\mathbbm{1}_{\binom{d}{n}} - 2 \sum_{j=1}^{\binom{d-2}{n-2}} \ket{j}\bra{j}~, &n>1~,
\end{array}
\right.
\end{equation}

and utilizing $\eta_{j,k} = \Pi_{i,j} \eta_{i,k} \Pi_{i,j}$ (see \cref{eq:RestBS}) the sign flips can be coped and all obtained commutation relations are conjectured to fulfill
\begin{subequations}
	\begin{align}
	\eta_{aj}[X_{a,j},Y_{b,k}]\eta_{aj} &= \eta_{bk}[X_{a,j},Y_{b,k}]\eta_{bk} = - \frac{i}{2} \delta_{a,b}X_{j,k},~&j\neq k~, 
	\label{eq:OffDiagsFermion1}\\
	\eta_{aj}[X_{a,j},X_{b,k}]\eta_{aj} &= \eta_{aj}[Y_{a,j},Y_{b,k}]\eta_{aj} = \eta_{bk}[X_{a,j},X_{b,k}]\eta_{bk} = \eta_{bk}[Y_{a,j},Y_{b,k}]\eta_{bk} =  \frac{i}{2}\delta_{a,b}Y_{j,k},~&j\neq k~.
	\label{eq:OffDiagsFermion2}
	\end{align}
\end{subequations}

For the diagonal matrices~\cref{eq:Diags} applies 

\begin{equation}
Z_{k}(n,d) \coloneqq -i\sqrt{\frac{2}{k(k+1)}}\sum_{j=1}^k j\, [X_{j,j+1}(n,d),Y_{j,j+1}(n,d)]~, \qquad 0<k<d~.
\label{eq:Diags}
\end{equation}

With \cref{eq:OffDiagsFermion1,eq:OffDiagsFermion2,eq:Diags} all algebraic generators can be linked to each other by commutation. When the off-diagonal generators of the algebra have no common lattice index ($a \neq b$) then they commute, else when they share one index~\cref{eq:OffDiagsFermion1,eq:OffDiagsFermion2} do not trivially vanish.
If both indices in the commutator are identical, the resulting matrix is diagonal and is either given by~\cref{eq:Diags} or by linear combinations thereof.
In the case of bosons or of a singular fermion $\eta$ can be omitted, but starting at 2 fermions in the 3-port MPI the bilinear expressions must be corrected.

\subsection{(Partially) differentiable modes}

Once the particles have observational differences, each of the particles show an individual behavior irrespective of their bosonic or fermionic nature. 
In order to find the mathematical description in terms of rotation matrices, previous ideas are now recycled. In the following lines an important restriction is made, namely the particles must not be different in any property related to their dispersion relation in the optical media.

\subsubsection{Full differentiability}

Treating each of two particles coincident on a single beam splitter as individual translates to 

\begin{equation}
e^{i \theta(\mathbbm{1}_2 \otimes S_y(2) + S_y(2) \otimes \mathbbm{1}_2)} = \left(
\begin{smallmatrix}
\cos ^2\frac{\theta }{2}& \cos \frac{\theta }{2} \sin \frac{\theta }{2} & \cos \frac{\theta }{2} \sin \frac{\theta }{2} & \sin ^2\frac{\theta }{2} \\
-\cos \frac{\theta }{2} \sin \frac{\theta }{2} & \cos ^2\frac{\theta }{2} & -\sin ^2\frac{\theta }{2} & \cos \frac{\theta }{2} \sin \frac{\theta }{2} \\
-\cos \frac{\theta }{2} \sin \frac{\theta }{2} & -\sin ^2\frac{\theta }{2} & \cos ^2\frac{\theta }{2} & \cos \frac{\theta }{2} \sin \frac{\theta }{2} \\
\sin ^2\frac{\theta }{2} & -\cos \frac{\theta }{2} \sin \frac{\theta }{2} & -\cos \frac{\theta }{2} \sin \frac{\theta }{2} & \cos ^2\frac{\theta }{2} \\
\end{smallmatrix}
\right)~.
\label{eq:Differential2particleBS}
\end{equation}

From the look of~\cref{eq:Differential2particleBS}, it is apparent that the which way interpretation of each entry is still valid. In contrast to the bosonic mode mixing in~\cref{eq:2Boson1BS} the $\{1,1\} \rightarrow \{1,1\}$ mode is now divided into two terms and the splitting $\{2\} \leftrightarrow \{1,1\}$ is no longer weighted by a $\sqrt{2}$ factor. Let us write the four input states as $\ket{\uparrow \downarrow, 0}, \ket{\uparrow, \downarrow}, \ket{ \downarrow, \uparrow}, \ket{0, \downarrow \uparrow}$. For a balanced beam splitter $\theta = \frac{\pi}{2}$ the transition probabilities are all $\frac{1}{4}$, meaning that for any of the four input state it is equally likely to find them in one of the 4 possible states. On the other hand, for the two Bell states

\begin{equation}
\ket{\psi^{\pm}} = \frac{1}{\sqrt{2}}(\ket{\uparrow, \downarrow} \pm \ket{ \downarrow, \uparrow}) = \frac{1}{\sqrt{2}} \left(0,1,\pm 1, 0\right)^T
\label{eq:BellStates}
\end{equation} 

the interference can lead to a HOM like effect $\ket{\psi^{+}} \rightarrow  \frac{1}{\sqrt{2}}(\ket{\uparrow \downarrow, 0} + \ket{0, \downarrow \uparrow})$,  where the output is a distinguishable N00N state, or not $\ket{\psi^{-}} \rightarrow  \ket{\psi^{-}}$, depending on the relative phase of the superposition. This argument holds for both types of particles, fermions and bosons.\\

Of course, we now seek to find a generalization to any number of particles or ports in the interferometer. For that purpose, the cyclic permutation of tensor products 

\begin{equation}
\pi (A_1 \otimes A_2 \otimes A_3 \otimes \cdots \otimes A_n) \coloneqq(A_n \otimes A_1 \otimes A_2 \otimes \cdots \otimes A_{n-1})
\end{equation}

is introduced. Applying the permutation $i$-times gives the $i$-fold cyclic permutation $\pi^i$ of the involved matrices. Looking back at~\cref{eq:Differential2particleBS} the operator in the exponent can be written as $\sum_{i=0}^{1} \pi^i(\mathbbm{1}_2 \otimes Y_{1,2}(1,2))$. More generally, for $n$-fully distinguishable particles in a $d$-sized MPI the beam splitters are described by

\begin{equation}
Y_{j,k}(nD,d)\coloneqq\sum_{i=0}^{n-1} \pi^i(\mathbbm{1}_{d}^{\otimes(n-1)} \otimes Y_{j,k}(1,d))~, \quad 1\le j<k \le d~.
\label{eq:MaximalDisting}
\end{equation}

Replacing the generators in~\cref{eq:LatticeUnitary} yields the desired description of the interferometer. It is worthwhile noting that $\exp(i \frac{\pi}{2} Y(nD,2))$ is, up to a different sign arrangement, equivalent to the Hadamard matrices of order $2^n$. Eventually, for the phase~\cref{eq:AllPS} can be recycled, as the shift is only dependent on the particle number.

\subsubsection{Partial differentiability}

For three or more particles it is possible to come up with a case like $\ket{\uparrow \uparrow \downarrow}$ where at least two of the constituents are identical but distinguish themselves from a third particle. It is not surprising that the shape of the generators is a mixture of the two extremes. In~\cref{eq:MaximalDisting}, the identity symbolizes an uninvolved system and the GGMs the individual particles. In the case of a singular beam splitter the given example of two out of three identical particles mathematically changes to

\begin{equation}\label{eq:ParitalSingleBS}
\begin{split}
Y_{1,2}(\ket{\uparrow \uparrow \downarrow}, 2)  &=\mathbbm{1}_{3} \otimes Y_{1,2}(1B,2) + Y_{1,2}(2B,2) \otimes \mathbbm{1}_{2}~,\qquad\text{(bosons)}~,\\
Y_{1,2}(\ket{\uparrow \uparrow \downarrow}, 2) &=\mathbbm{1}_1 \otimes Y_{1,2}(1F,2) + Y_{1,2}(2F,2) \otimes \mathbbm{1}_2~,\qquad\text{(fermions)}~.
\end{split}
\end{equation}

where the identical $\ket{\uparrow \uparrow}$ - state is determined by the term $\{\mathbbm{1}_{3}, Y_{1,2}(2B,2)\}$ and the single $\ket{\downarrow}$ by $\{\mathbbm{1}_{2},Y_{1,2}(1,2)\}$ for bosons (the argument goes similarly for fermions). 
Expanding the example given to the interferometer with three ports results in

\begin{equation}\label{eq:partiallydist23}
\begin{split}
Y_{j,k}(\ket{\uparrow \uparrow \downarrow}, 3) &= \mathbbm{1}_6 \otimes Y_{j,k}(1B,3) + Y_{j,k}(2B,3) \otimes \mathbbm{1}_3~,\qquad\text{(bosons)}~,\\
Y_{j,k}(\ket{\uparrow \uparrow \downarrow}, 3) &=\mathbbm{1}_3 \otimes Y_{j,k}(1F,3) + Y_{j,k}(2F,3) \otimes \mathbbm{1}_3~,\qquad\text{(fermions)}~.
\end{split}
\end{equation}

\cref{eq:partiallydist23} highlights that in the case of partially distinguishable systems, the generators of the rotations can be assembled modularly. This eliminates the need for a separate analysis of this case. It should be mentioned that a simulation of distinguishable particles with a single particle, similar to~\cref{tab:Wigner1DTable}, is once more possible.

\section{Discussion}

This manuscript provides a time-independent model for describing idealized multi-port interferometers for single and multiple particles. The universal method is based on finding the rotation matrices, obtained by exponential mapping of the associated algebraic generators, and multiplying the elements in the specific order as in~\cref{eq:LatticeUnitary}. In the case of a single particle, a recursive approach scheme allows for solutions of the form~\cref{eq:Solution}.
For more than one particle the distinction to different cases is necessary. It is shown here how to find the right operators to set up the entire unitary transition matrix. The proposed construction scheme can be used for an explicit representation both in the bosonic and fermionic case~\cref{eq:LastBS,eq:RestBS,eq:AllPS}. Finally, it has been demonstrated that beam splitters for distinguishable particles have a convenient representation~\cref{eq:MaximalDisting} by tensor products.\\

The concepts presented here invite to many generalizations and further studies. Some interesting ideas are given in the following lines.

\begin{itemize}
	\item All unitary operators presented here are particle number preserving. When incorporating non-linear, active optical elements it is possible to device MPIs which alter the particle number. An interesting class is given by the set of $SU(m,n)$ interferometers~\cite{Yurke1986}.
	\item It is possible to mix the concepts with coherent states~\cite{Windhager2011} and analyze the continuous limit (path or particle number $(d,n) \rightarrow \infty$). By determining the associated Haar measure it is likewise possible \cite{Spengler2019} to obtain the infinitesimal volume element $dU^{lat}(n, d)$.
	\item Considering $\epsilon (\rho) = U\rho U^{\dagger}$ as a unitary quantum channel of the quantum state $\rho$ by the unitary lattice operator~\cref{eq:LatticeUnitary}, it is compelling to extend the concept to trace non-increasing quantum operations, that also take losses or decohering effects into consideration. Especially investigating a controlled mapping to the space of high dimensional mixed states would be interesting.
	\item There are many basic variations of multi-interference designs; they may differ in the architecture of the spatial modes~\cite{Clements16}, exchange spatial with temporal overlaps~\cite{Motes2014}, rotation~\cite{Ivanov2006} for reflections or increase the number of paths at a single node~\cite{Weihs1996}. The comparison and applicability of the results shown here for other cases would be interesting.
\end{itemize}
\newpage

\iffalse
\section*{Acknowledgment}
Thanks
\fi

\begin{appendices}

\section{Appendix}
\crefalias{subsection}{app}

\begin{figure}[h!]
	\centering
	\subfloat[][\centering4-port interferometer with $R_{1,4} = 1$ and $R_{1,2} = 0$ and .]{\includegraphics[width=0.4\textwidth]{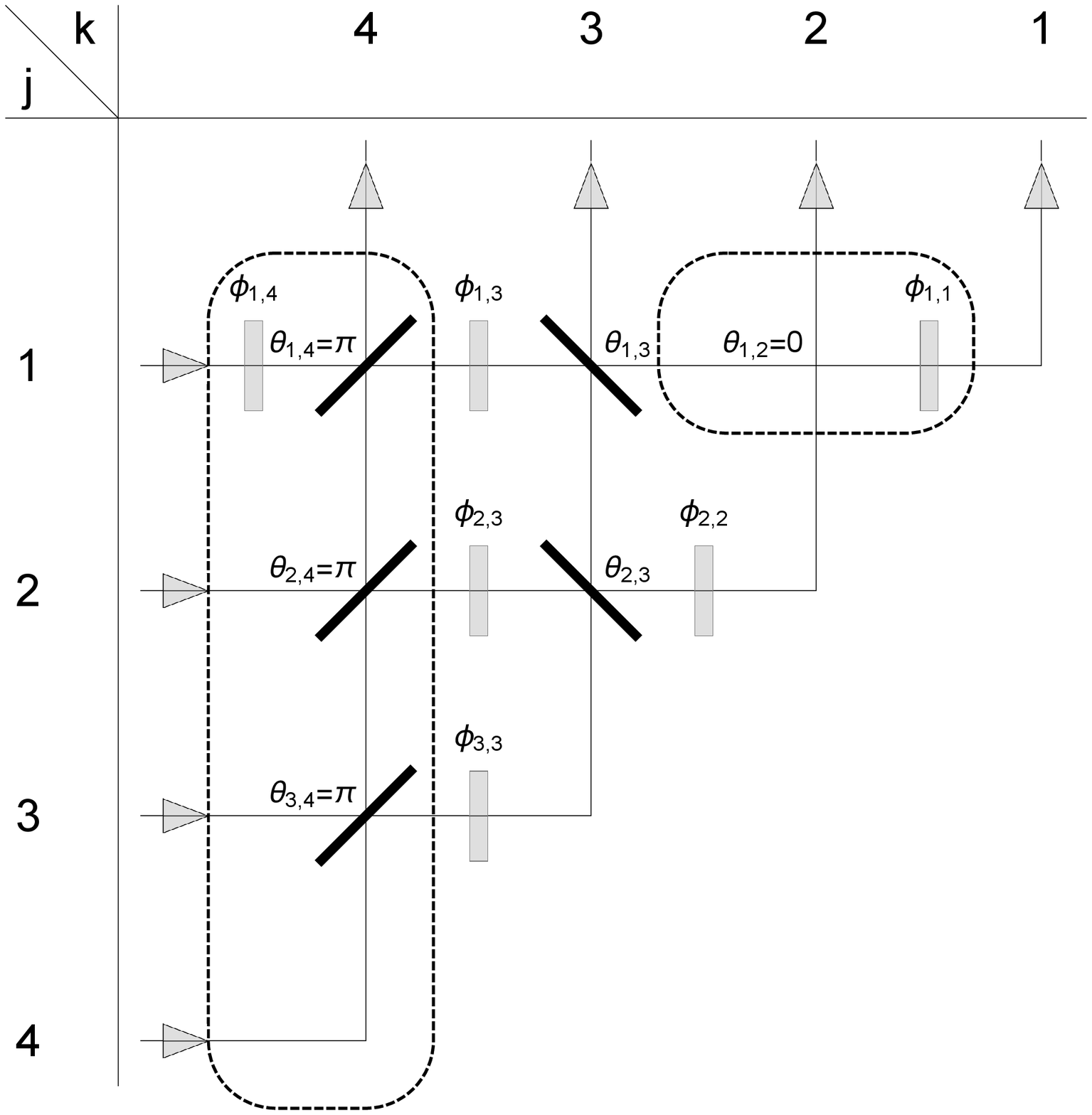}\label{fig:4Portdef1}}
	\hspace{7mm}
	\subfloat[][\centering Decay of the interferometer into two parts. .]{\raisebox{10mm}{\includegraphics[width=0.4\textwidth]{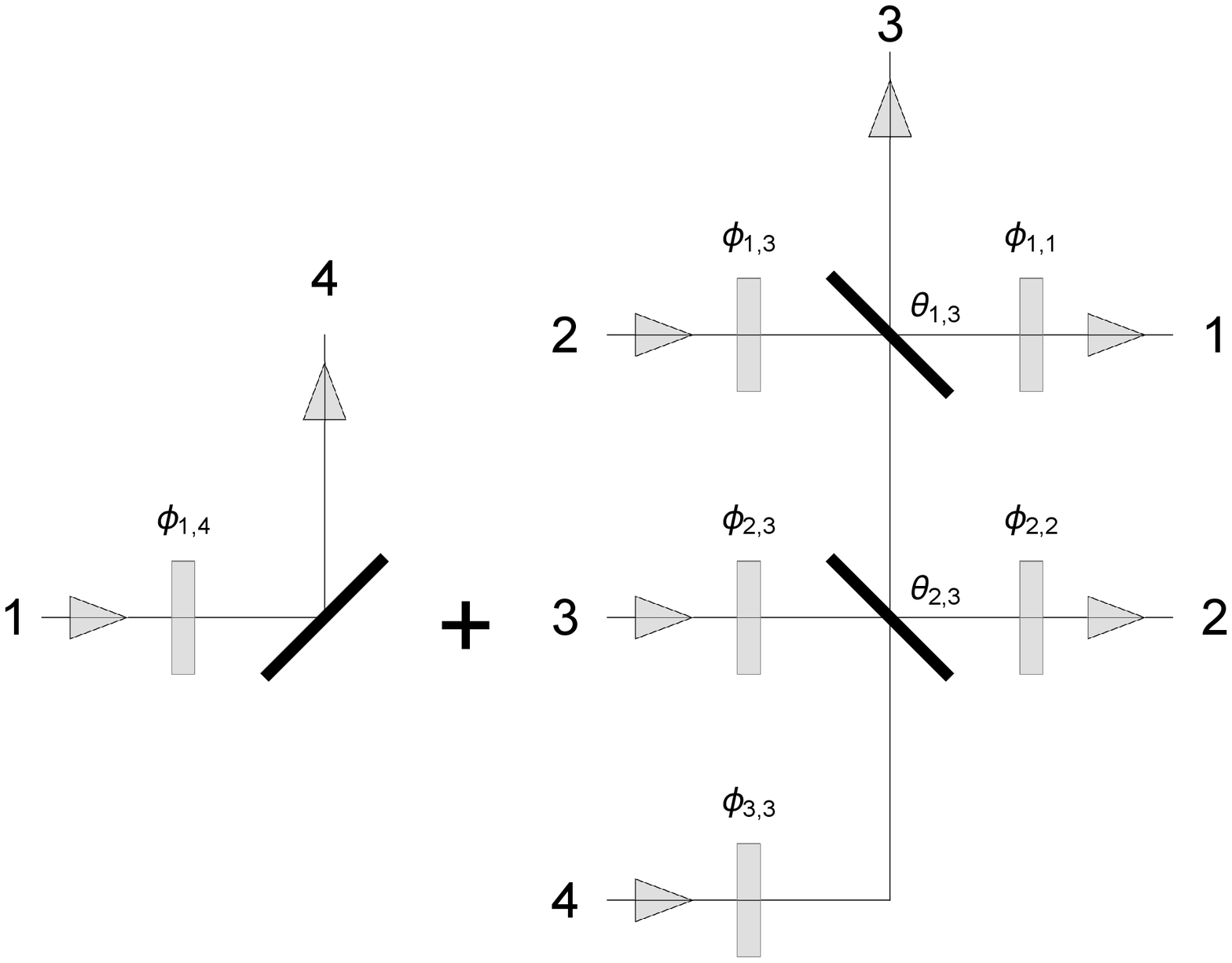}}\label{fig:4Portdef2}}
	\caption{(a) On the left~\cref{fig:4Portdef1} the two extreme cases are sketched. In the larger dashed box an obvious problem occurs when one node $R_{1,4} = 1$ is totally reflective, since it completely blocks input port~1, respectively output port~4. As a consequence all beam splitters below that node are set to be totally reflective as well and all phase shifter with the same output index are set to zero, hence they are no longer drawn. Since e.g. $R_{1,3}$ cannot be determined from the equation obtained when going from $j=1$ to $k=3$ the integer $b$ in~\cref{eq:NomDenom2,eq:DefintionOfZ2} is increased by one to jump to the next row $j=2$. The smaller dashed box of the figure indicates that the phase shift is fixed $\phi_{j,k} =0$ whenever the node is inactive. (b) The right side of~\cref{fig:4Portdef2} illustrates the decomposition of the MPI to it's left into two parts.}
	\label{fig:4Portdefs}
\end{figure}

\subsection{Appendix: Extreme cases}

\label{app:Appendix1}
From the look at the formulae in~\cref{eq:Solution} it is clear that there are two possible issues that need to be handled. The lesser problem is given by 

\begin{itemize}
	\item[--] \textit{Total transmission:} $R_{j,k} = Z_{j,k} = 0$. When the probability for reflection at node $(j,k)$ is zero, there is maximal transmission without interference at the intersection. In that case a succession of two phase shifters in the horizontal paths occurs. To avoid an over-determination in the parameter space of the lattice operator it is the best to set $x_{j,k}=1$, which implies that the phase shift is zero.
\end{itemize}

A more inconvenient situation arises for

\begin{itemize}
	\item[--] \textit{Total reflection:} $R_{jk} = 1$. When at some point the transmission probability is 0, the denominator $D_{jk}$ in~\cref{eq:Denom} eventually turns to 0 and all paths from the $j+1$-th input to the $k$-th output become blocked. For the algorithm to be still working, it is the best to set the reflectivity of all nodes below the affected one, i.e. $\{(j+1,k),(j+2,k),(j+\ldots,k)\}$, to maximum. Consequently, there is again an abundance of phase shifters, hence $\phi_{j,k} = 0$ for all $j$'s after which total reflectivity occurred. The situation is shown in~\cref{fig:4Portdefs}. 
	
	It is also apparent that other probabilities, e.g. $R_{1,3}$, can no longer be determined from the path leading from the 1st input to the 3rd output port. Instead it is necessary to jump into the succeeding 2nd beam path which in terms of the lattice matrix~\cref{eq:LatticeUnitary} means that its necessary to go to the next possible row for determination. In~\cref{eq:NomDenom2} this is taken care of by parameter \textit{b} which increases whenever a jump to the lower row occurs.
\end{itemize}

\begin{equation}
\begin{array}{l}
N_{j,k} \coloneqq \sum\limits_{m=1}^{j-1}e^{i \phi_{m,k}}\sqrt{R_{m,k}}\braket{j+b|e(m,k)}\prod\limits_{l=1}^{m-1}\sqrt{1-R_{l,k}} \\[6pt]
D_{j,k} \coloneqq \braket{j+b|e(j,k)}\prod\limits_{l=1}^{j-1}\sqrt{1-R_{l,k}}
\end{array}
\label{eq:NomDenom2}
\end{equation}
and
\begin{equation}
Z_{j,k} \coloneqq \frac{U^{tar}_{j+b,k} - N_{j,k}}{D_{j,k}}
\label{eq:DefintionOfZ2}
\end{equation}

Here, the integer $b$ counts how often the beam path from \textit{k} to \textit{j} is blocked and is initialized as 0.

\subsection{Calculation of rotation matrices - Examples in the manuscript}
\label{app:Appendix2}

\subsubsection{MPI for n=2B, d=3}

Step 1.
\begin{equation*}
\begin{split}
\ket{\text{co}(2B,3)} &= \left\{\ket{2,0,0}, \ket{1,1,0}, \ket{1,0,1}, \ket{0,2,0}, \ket{0,1,1}, \ket{0,0,2} \right\}~,\\ SW_{1,2}\ket{\text{co}(2,3)} &= \left\{\ket{0,2,0}, \ket{1,1,0}, \ket{0,1,1}, \ket{2,0,0}, \ket{1,0,1},\ket{0,0,2} \right\}~,\\ SW_{1,3}\ket{\text{co}(2,3)} &= \left\{\ket{0,0,2}, \ket{0,1,1}, \ket{1,0,1}, \ket{0,2,0}, \ket{1,1,0}, \ket{2,0,0} \right\}~,\\ SW_{2,3}\ket{\text{co}(2,3)} &= \left\{\ket{2,0,0}, \ket{1,0,1}, \ket{1,1,0}, \ket{0,0,2}, \ket{0,1,1}, \ket{0,2,0} \right\}~.
\end{split}
\end{equation*}

$$P_i = \{P\left(\bra{\text{co}(2B,3)}  N_3 \ket{\text{co}(2B,3)}\right)\} = \{\{0\}, \{0,1\}, \{0,1,2\}\}$$

Step 2.
\begin{equation*}
\begin{split}
\Pi_{1,2} = \ket{0,2,0}\bra{2,0,0}+ \ket{1,1,0}\bra{1,1,0}+ \ket{0,1,1}\bra{1,0,1}+ \ket{2,0,0}\bra{0,2,0}+ \ket{1,0,1} \bra{0,1,1}+\ket{0,0,2}\bra{0,0,2}~,\\ \Pi_{1,3} = \ket{0,0,2}\bra{2,0,0}+ \ket{0,1,1}\bra{1,1,0}+ \ket{1,0,1}\bra{1,0,1}+ \ket{0,2,0}\bra{0,2,0}+ \ket{1,1,0} \bra{0,1,1}+\ket{2,0,0}\bra{0,0,2}~,\\ \Pi_{2,3} = \ket{2,0,0}\bra{2,0,0}+ \ket{1,0,1}\bra{1,1,0}+ \ket{1,1,0}\bra{1,0,1}+ \ket{0,0,2}\bra{0,2,0}+ \ket{0,1,1} \bra{0,1,1}+\ket{0,2,0}\bra{0,0,2}~.
\end{split}
\end{equation*}

Step 3. Beam splitter:

$$Y_{2,3}(2B,3) = S_y(1) \oplus S_y(2) \oplus S_y(3) = \left(
\begin{smallmatrix}
0 & 0 & 0 & 0 & 0 & 0 \\0 & 0 & -\frac{i}{2} & 0 & 0 & 0 \\0 & \frac{i}{2} & 0 & 0 & 0 & 0 \\
0 & 0 & 0 & 0 & -\frac{i}{\sqrt{2}} & 0 \\0 & 0 & 0 & \frac{i}{\sqrt{2}} & 0 & -\frac{i}{\sqrt{2}} \\
0 & 0 & 0 & 0 & \frac{i}{\sqrt{2}} & 0 \\
\end{smallmatrix}
\right)$$

$$Y_{1,3}(2B,3) = \Pi_{1,2} Y_{2,3} \Pi_{1,2} =  \left(
\begin{smallmatrix}0 & 0 & 0 & 1 & 0 & 0 \\0 & 1 & 0 & 0 & 0 & 0 \\0 & 0 & 0 & 0 & 1 & 0 \\
1 & 0 & 0 & 0 & 0 & 0 \\0 & 0 & 1 & 0 & 0 & 0 \\0 & 0 & 0 & 0 & 0 & 1 \\\end{smallmatrix}\right)
\left(\begin{smallmatrix}
0 & 0 & 0 & 0 & 0 & 0 \\0 & 0 & -\frac{i}{2} & 0 & 0 & 0 \\0 & \frac{i}{2} & 0 & 0 & 0 & 0 \\
0 & 0 & 0 & 0 & -\frac{i}{\sqrt{2}} & 0 \\0 & 0 & 0 & \frac{i}{\sqrt{2}} & 0 & -\frac{i}{\sqrt{2}} \\
0 & 0 & 0 & 0 & \frac{i}{\sqrt{2}} & 0 \\
\end{smallmatrix}
\right)\left(
\begin{smallmatrix}0 & 0 & 0 & 1 & 0 & 0 \\0 & 1 & 0 & 0 & 0 & 0 \\0 & 0 & 0 & 0 & 1 & 0 \\
1 & 0 & 0 & 0 & 0 & 0 \\0 & 0 & 1 & 0 & 0 & 0 \\0 & 0 & 0 & 0 & 0 & 1 \\\end{smallmatrix}\right) = \left(
\begin{smallmatrix}0 & 0 & -\frac{i}{\sqrt{2}} & 0 & 0 & 0 \\0 & 0 & 0 & 0 & -\frac{i}{2} & 0 \\
\frac{i}{\sqrt{2}} & 0 & 0 & 0 & 0 & -\frac{i}{\sqrt{2}} \\0 & 0 & 0 & 0 & 0 & 0 \\
0 & \frac{i}{2} & 0 & 0 & 0 & 0 \\0 & 0 & \frac{i}{\sqrt{2}} & 0 & 0 & 0 \\
\end{smallmatrix}
\right)$$

$$Y_{1,2}(2B,3) = -\Pi_{1,3} Y_{2,3} \Pi_{1,3} = \left(
\begin{smallmatrix}0 & 0 & 0 & 0 & 0 & 1 \\0 & 0 & 0 & 0 & 1 & 0 \\0 & 0 & 1 & 0 & 0 & 0 \\
0 & 0 & 0 & 1 & 0 & 0 \\0 & 1 & 0 & 0 & 0 & 0 \\1 & 0 & 0 & 0 & 0 & 0 \\
\end{smallmatrix}\right)
\left(\begin{smallmatrix}
0 & 0 & 0 & 0 & 0 & 0 \\0 & 0 & -\frac{i}{2} & 0 & 0 & 0 \\0 & \frac{i}{2} & 0 & 0 & 0 & 0 \\
0 & 0 & 0 & 0 & -\frac{i}{\sqrt{2}} & 0 \\0 & 0 & 0 & \frac{i}{\sqrt{2}} & 0 & -\frac{i}{\sqrt{2}} \\
0 & 0 & 0 & 0 & \frac{i}{\sqrt{2}} & 0 \\
\end{smallmatrix}
\right)\left(
\begin{smallmatrix}0 & 0 & 0 & 1 & 0 & 0 \\0 & 1 & 0 & 0 & 0 & 0 \\0 & 0 & 0 & 0 & 1 & 0 \\
1 & 0 & 0 & 0 & 0 & 0 \\0 & 0 & 1 & 0 & 0 & 0 \\0 & 0 & 0 & 0 & 0 & 1 \\\end{smallmatrix}\right)= \left(
\begin{smallmatrix}
0 & -\frac{i}{\sqrt{2}} & 0 & 0 & 0 & 0 \\
\frac{i}{\sqrt{2}} & 0 & 0 & -\frac{i}{\sqrt{2}} & 0 & 0 \\
0 & 0 & 0 & 0 & -\frac{i}{2} & 0 \\
0 & \frac{i}{\sqrt{2}} & 0 & 0 & 0 & 0 \\
0 & 0 & \frac{i}{2} & 0 & 0 & 0 \\
0 & 0 & 0 & 0 & 0 & 0 \\
\end{smallmatrix}
\right)$$

Phase shifter:

\begin{equation*}
\begin{split}
E_{1}(2B,3) = \text{diag}(2,1,1,0,0,0)~,\\ E_{2}(2B,3) = \text{diag}(0,1,0,2,1,0)~.\\ E_{3}(2B,3) = \text{diag}(0,0,1,0,1,2)~.
\end{split}
\end{equation*}

\subsubsection{MPI for n=2F, d=4}

Step 1.
\begin{equation*}
\begin{split}
\ket{\text{co}(2F,4)} &= \left\{\ket{1,1,0,0}, \ket{1,0,1,0}, \ket{1,0,0,1}, \ket{0,1,1,0}, \ket{0,1,0,1}, \ket{0,0,1,1} \right\}~,\\ SW_{1,2}\ket{\text{co}(2F,4)} &= \left\{\ket{1,1,0,0}, \ket{0,1,1,0}, \ket{0,1,0,1}, \ket{1,0,1,0}, \ket{1,0,0,1}, \ket{0,0,1,1} \right\}~,\\ SW_{1,3}\ket{\text{co}(2F,4)} &= \left\{\ket{0,1,1,0}, \ket{1,0,1,0}, \ket{0,0,1,1}, \ket{1,1,0,0}, \ket{0,1,0,1}, \ket{1,0,0,1} \right\}~,\\ SW_{1,4}\ket{\text{co}(2F,4)} &= \left\{\ket{0,1,0,1}, \ket{0,0,1,1}, \ket{1,0,0,1}, \ket{0,1,1,0}, \ket{1,1,0,0}, \ket{1,0,1,0} \right\}~,\\SW_{2,3}\ket{\text{co}(2F,4)} &= \left\{\ket{1,0,1,0}, \ket{1,1,0,0}, \ket{1,0,0,1}, \ket{0,1,1,0}, \ket{0,0,1,1}, \ket{0,1,0,1} \right\}~,\\ SW_{2,4}\ket{\text{co}(2F,4)} &= \left\{\ket{1,0,0,1}, \ket{1,0,1,0}, \ket{1,1,0,0}, \ket{0,0,1,1}, \ket{0,1,0,1}, \ket{0,1,1,0} \right\}~,\\ SW_{3,4}\ket{\text{co}(2F,4)} &= \left\{\ket{1,1,0,0}, \ket{1,0,0,1}, \ket{1,0,1,0}, \ket{0,1,0,1}, \ket{0,1,1,0}, \ket{0,0,1,1} \right\}~.
\end{split}
\end{equation*}

$$P_i = \{P \left(\bra{\text{co}(2F,4)}  N_4 \ket{\text{co}(2F,4)}\right)\} = \{\{0\}, \{0,1\}, \{0,1\}, \{1\}\}~.$$

Step 2.
\begin{gather*}
\Pi_{1,2} = \resizebox{.9\textwidth}{!}{$\ket{1,1,0,0}\bra{1,1,0,0}+ \ket{1,0,1,0}\bra{0,1,1,0}+ \ket{1,0,0,1}\bra{0,1,0,1}+ \ket{0,1,1,0}\bra{1,0,1,0}+ \ket{0,1,0,1} \bra{1,0,0,1}+\ket{0,0,1,1}\bra{0,0,1,1}$}~,\\ \Pi_{1,3} = \resizebox{.9\textwidth}{!}{$\ket{1,1,0,0}\bra{0,1,1,0}+ \ket{1,0,1,0}\bra{1,0,1,0}+ \ket{1,0,0,1}\bra{0,0,1,1}+ \ket{0,1,1,0}\bra{1,1,0,0}+ \ket{0,1,0,1} \bra{0,1,0,1}+\ket{0,0,1,1}\bra{1,0,0,1}$}~,\\\Pi_{1,4} = \resizebox{.9\textwidth}{!}{$\ket{1,1,0,0}\bra{0,1,0,1}+ \ket{1,0,1,0}\bra{0,0,1,1}+ \ket{1,0,0,1}\bra{1,0,0,1}+ \ket{0,1,1,0}\bra{0,1,1,0}+ \ket{0,1,0,1} \bra{1,1,0,0}+\ket{0,0,1,1}\bra{1,0,1,0}$}~,\\\Pi_{2,3} = \resizebox{.9\textwidth}{!}{$\ket{1,1,0,0}\bra{1,0,1,0}+ \ket{1,0,1,0}\bra{1,1,0,0}+ \ket{1,0,0,1}\bra{1,0,0,1}+ \ket{0,1,1,0}\bra{0,1,1,0}+ \ket{0,1,0,1} \bra{0,0,1,1}+\ket{0,0,1,1}\bra{0,1,0,1}$}~,\\\Pi_{2,4} = \resizebox{.9\textwidth}{!}{$\ket{1,1,0,0}\bra{1,0,0,1}+ \ket{1,0,1,0}\bra{1,0,1,0}+ \ket{1,0,0,1}\bra{1,1,0,0}+ \ket{0,1,1,0}\bra{0,0,1,1}+ \ket{0,1,0,1} \bra{0,1,0,1}+\ket{0,0,1,1}\bra{0,1,1,0}$}~,\\\Pi_{3,4} = \resizebox{.9\textwidth}{!}{$\ket{1,1,0,0}\bra{1,1,0,0}+ \ket{1,0,1,0}\bra{1,0,0,1}+ \ket{1,0,0,1}\bra{1,0,1,0}+ \ket{0,1,1,0}\bra{0,1,0,1}+ \ket{0,1,0,1} \bra{0,1,1,0}+\ket{0,0,1,1}\bra{0,0,1,1}$}~.
\end{gather*}

Step 3. Beam splitter:

\begin{flalign*}
Y_{3,4}(2F,4) &= S_y(1) \oplus S_y(2) \oplus S_y(2) \oplus S_y(1) = \left(
\begin{smallmatrix}
0 & 0 & 0 & 0 & 0 & 0 \\
0 & 0 & -\frac{i}{2} & 0 & 0 & 0 \\
0 & \frac{i}{2} & 0 & 0 & 0 & 0 \\
0 & 0 & 0 & 0 & -\frac{i}{2} & 0 \\
0 & 0 & 0 & \frac{i}{2} & 0 & 0 \\
0 & 0 & 0 & 0 & 0 & 0 \\
\end{smallmatrix}
\right)~,\\
Y_{2,4}(2F,4) &= \Pi_{2,3} Y_{3,4} \Pi_{2,3} = \left(
\begin{smallmatrix}
0 & 0 & -\frac{i}{2} & 0 & 0 & 0 \\
0 & 0 & 0 & 0 & 0 & 0 \\
\frac{i}{2} & 0 & 0 & 0 & 0 & 0 \\
0 & 0 & 0 & 0 & 0 & -\frac{i}{2} \\
0 & 0 & 0 & 0 & 0 & 0 \\
0 & 0 & 0 & \frac{i}{2} & 0 & 0 \\
\end{smallmatrix}
\right)~,\\
Y_{1,4}(2F,4) &= \Pi_{1,3} Y_{3,4} \Pi_{1,3} = \Pi_{1,2} Y_{2,4} \Pi_{1,2} = \left(
\begin{smallmatrix}
0 & 0 & 0 & 0 & -\frac{i}{2} & 0 \\
0 & 0 & 0 & 0 & 0 & -\frac{i}{2} \\
0 & 0 & 0 & 0 & 0 & 0 \\
0 & 0 & 0 & 0 & 0 & 0 \\
\frac{i}{2} & 0 & 0 & 0 & 0 & 0 \\
0 & \frac{i}{2} & 0 & 0 & 0 & 0 \\
\end{smallmatrix}
\right)~,\\
Y_{2,3}(2F,4) &= -\Pi_{2,4} Y_{3,4} \Pi_{2,4} = \Pi_{3,4} Y_{2,4} \Pi_{3,4} = \left(
\begin{smallmatrix}
0 & -\frac{i}{2} & 0 & 0 & 0 & 0 \\
\frac{i}{2} & 0 & 0 & 0 & 0 & 0 \\
0 & 0 & 0 & 0 & 0 & 0 \\
0 & 0 & 0 & 0 & 0 & 0 \\
0 & 0 & 0 & 0 & 0 & -\frac{i}{2} \\
0 & 0 & 0 & 0 & \frac{i}{2} & 0 \\
\end{smallmatrix}
\right)~,\\
Y_{1,3}(2F,4) &= -\Pi_{1,4} Y_{3,4} \Pi_{1,4} = \Pi_{3,4} Y_{1,4} \Pi_{3,4} = \Pi_{1,2} Y_{2,3} \Pi_{1,2} = \left(
\begin{smallmatrix}
0 & 0 & 0 & -\frac{i}{2} & 0 & 0 \\
0 & 0 & 0 & 0 & 0 & 0 \\
0 & 0 & 0 & 0 & 0 & -\frac{i}{2} \\
\frac{i}{2} & 0 & 0 & 0 & 0 & 0 \\
0 & 0 & 0 & 0 & 0 & 0 \\
0 & 0 & \frac{i}{2} & 0 & 0 & 0 \\
\end{smallmatrix}
\right)~,\\
Y_{1,2}(2F,4) &= -\Pi_{1,4} Y_{2,4} \Pi_{1,4} = \Pi_{2,4} Y_{1,4} \Pi_{2,4} = -\Pi_{1,3} Y_{2,3} \Pi_{1,3} = \Pi_{2,3} Y_{1,3} \Pi_{2,3} = \left(\begin{smallmatrix}
0 & 0 & 0 & 0 & 0 & 0 \\
0 & 0 & 0 & -\frac{i}{2} & 0 & 0 \\
0 & 0 & 0 & 0 & -\frac{i}{2} & 0 \\
0 & \frac{i}{2} & 0 & 0 & 0 & 0 \\
0 & 0 & \frac{i}{2} & 0 & 0 & 0 \\
0 & 0 & 0 & 0 & 0 & 0 \\
\end{smallmatrix}
\right)~. \\
\end{flalign*}

Phase shifter:

\begin{equation*}
\begin{split}
E_{1}(2F,4) = \text{diag}(1,1,1,0,0,0)~,\qquad E_{2}(2F,4) = \text{diag}(1,0,0,1,1,0)~,\\ E_{3}(2F,4) = \text{diag}(0,1,0,1,0,1)~,\qquad E_{4}(2F,4) = \text{diag}(0,0,1,0,1,1)~.
\end{split}
\end{equation*}

\end{appendices}

\section{Bibliography}

\bibliographystyle{apsrev4-1}
\bibliography{mybibfileMPI}

\end{document}